\documentclass[pdflatex,sn-mathphys-num]{sn-jnl}
\usepackage[tablesfirst, notablist, figlist, disable]{endfloat}
\usepackage[caption=false]{subfig}
\usepackage[title]{appendix}%
\usepackage{amssymb}
\usepackage{siunitx}
\usepackage{mathtools}
\usepackage{multirow}
\usepackage{colortbl}
\usepackage{enumitem}
\usepackage{xcolor}
\usepackage{booktabs} 
\usepackage{placeins}
\usepackage{subfig}
\usepackage{tikz}
\usetikzlibrary{tikzmark}
\usepackage[english]{babel}
\usepackage{multicol}
\usepackage[normalem]{ulem}
\usepackage{todonotes}
\usepackage{acronym}
\acrodef{MOR}{Model Order Reduction}
\acrodef{PMOR}{Parametric Model Order Reduction}
\acrodef{EB}{Euler--Bernoulli}
\acrodef{ODE}{Ordinary Differential Equation}
\acrodef{DAE}{Differential Algebraic Equation}
\acrodef{DOF}{Degree of Freedom}
\acrodef{BC}{Boundary Condition}
\acrodef{RMS}{Root-Mean-Square}
\acrodef{POD}{Proper Orthogonal Decomposition}
\acrodef{MBS}{MultiBody System}
\acrodef{MBD}{MultiBody Dynamics}
\acrodefplural{DOF}[DOFs]{Degrees of Freedom}
\acrodef{ANCF}{Absolute Nodal Coordinate Formulation}
\acrodef{SVD}{Singular Value Decomposition}

\newcommand{\eqdot}{\, .} 
\newcommand{\eqcomma}{\, ,} 

\newcommand{\vtretp}[3]{\left[ \begin{array}{ccc} #1 & #2 & #3 \end{array} \right]}
\newcommand{\vfoutp}[4]{\left[ \begin{array}{cccc} #1 & #2 & #3 & #4 \end{array} \right]}

\newcommand{\ev}{\mathbf{e}}
\newcommand{\fv}{\mathbf{f}}
\newcommand{\gv}{\mathbf{g}}

\newcommand{\pv}{\mathbf{p}}
\newcommand{\qv}{\mathbf{q}}
\newcommand{\rv}{\mathbf{r}}
\newcommand{\sv}{\mathbf{s}}

\newcommand{\Cm}{\mathbf{C}}
\newcommand{\Km}{\mathbf{K}}

\newcommand{\Sm}{\mathbf{S}}
\newcommand{\Mm}{\mathbf{M}}

\newcommand{\Um}{\mathbf{U}}
\newcommand{\Vm}{\mathbf{V}}

\newcommand{\tPsi}{\mbox{$\pmb{\Psi\;}$}\!\!}
\newcommand{\tSigma}{\mbox{$\pmb{\Sigma\;}$}\!\!}
\newcommand{\tlambda}{\mbox{$\pmb{\lambda}\;$}\!\!}
\newcommand{\tvarphi}{\mbox{$\pmb{\varphi\;}$}\!\!}
\newcommand{\tp}{^\mathrm{T}}
\newcommand{\ANCF}{^\mathrm{ANCF}}
\newcommand{\ele}{_\mathrm{e}}
\newcommand{\nod}{_\mathrm{n}}
\newcommand{\nDOFs}{n_\mathrm{DOF}}
\newcommand{\mult}{\,}
\newcommand{\snap}{_\mathrm{snap}}
\newcommand{\glob}{_\mathrm{glob}}
\newcommand{\red}{_\mathrm{red}}
\newcommand{\nr}{k}
\newcommand{\Null}{\mathbf{0}}
\newcommand{\One}{\mathbf{1}}

\newcommand{\EXU}{Exudyn}
\usepackage[most]{tcolorbox}
\newtcolorbox{todobox}[1][]{
  colback=yellow!10,
  colframe=orange!70!black,
  title=Writing Task\ifx&#1&\else{} #1\fi,
  fonttitle=\bfseries,
  arc=3mm,
  boxrule=0.8pt,
  left=2mm,
  right=2mm,
  top=1mm,
  bottom=1mm
}

\begin{document}
\title[BoomPMR]{Model Order Reduction of a Sliding Beam using a Global Basis: Formulation and Evaluation}
\author[1,2]{\fnm{Sebastian} \sur{Weyrer}}\email{sebastian.weyrer@uibk.ac.at}
\author[2]{\fnm{Johannes} \sur{Gerstmayr}}\email{johannes.gerstmayr@uibk.ac.at}
\author[1]{\fnm{Aki} \sur{Mikkola}}\email{aki.mikkola@lut.fi}
\author*[1]{\fnm{Grzegorz} \sur{Orzechowski}}\email{grzegorz.orzechowski@lut.fi}
\affil[1]{\orgdiv{Department of Mechanical Engineering}, \orgname{Lappeenranta-Lahti University of Technology}, \orgaddress{\city{Lappeenranta and Lahti}, \country{Finland}}}
\affil[2]{\orgdiv{Department of Mechatronics}, \orgname{University of Innsbruck}, \orgaddress{\street{Technikerstraße 13}, \city{Innsbruck}, \postcode{6020}, \state{Tyrol}, \country{Austria}}}

\abstract{Model order reduction decreases the dimension of a mechanical system by introducing modal coordinates that retain important dynamic characteristics. Sliding beams, as found in telescopic structures, pose a fundamental challenge. Fixed modal coordinates fail to capture evolving system properties, and updating the modal basis during simulation causes modal coordinates to change meaning. The present work addresses this challenge by constructing a global reduction basis for a sliding beam. The global basis is constructed from snapshots in the form of modal matrices and compressed using proper orthogonal decomposition. Reduction is applied within a constraint multibody formalism with algebraically enforced constraints that permit continuous slider movement. The method is validated against an absolute nodal coordinate formulation of a sliding beam with a sliding joint. Different combinations of snapshot quantity and eigenmodes per snapshot are investigated and an error map is shown. A challenging test case involving a highly flexible beam subjected to time-dependent loading and slider movement demonstrates that the global reduction basis reduces computation time by approximately 90\% while keeping the root-mean-square displacement error, introduced by the global reduction, below 2\%.}
\maketitle

\section{Introduction}
\label{Introduction}
Telescopic structures are a fundamental component of heavy machinery. They are used in cranes, aerial work platforms, telehandlers, and forestry machines, where adjustable reach is an operational requirement. Telescopic booms enable machines to vary their working radius from a compact transport configuration to full extension. At the same time, they present a demanding challenge. As a telescoping boom expands, its structure becomes increasingly flexible, and managing dynamic loads requires a thorough understanding of structural behaviors throughout the range of motion.

The sliding beams inherent in telescopic booms are defining mechanical features because their effective lengths and \acp{BC} change continuously. These parameter-dependent topology changes make modeling and simulation particularly challenging, an issue addressed, for example, by Steinbrecher et al.~\cite{Steinbrecher2017} and Humer et al.~\cite{Humer2020}. However, the computational cost of high-fidelity discretizations remains an obstacle, especially when simulations must be performed repeatedly in the context of design optimization or real-time applications.

\ac{MOR} offers a well-established remedy for large-scale discretized flexible systems. The foundational works of Hurty~\cite{Hurty1965} and Craig and Bampton~\cite{Craig1968} demonstrate that a structure's dynamic behavior can be captured accurately using a small set of modal coordinates. However, classical reduction techniques assume fixed topologies: the mode shapes are computed once and remain valid throughout the simulation. This assumption breaks down for sliding beams, where effective lengths, \acp{BC}, and mode shapes change continuously as beams extend or retract. A straightforward application of classical reduction would require recomputing and replacing modal coordinates every time these parameters change. This effectively defeats the purpose of the reduction.

\ac{PMOR} provides a systematic framework to address this difficulty. \ac{PMOR} methods can be broadly divided into local approaches, which interpolate between reduced models computed at discrete parameter values, and global approaches, which construct a single reduction basis valid across an entire parameter range. Comprehensive overviews are provided by Benner et al.~\cite{Benner2015}, a survey that has become a standard reference in \ac{PMOR}, and Panzer et al.~\cite{Panzer2010}, who also introduce a local method based on interpolating locally reduced system matrices after aligning coordinate systems. Amsallem and Farhat~\cite{Amsallem2008, Amsallem2011} propose an alternative local approach that maps precomputed bases onto a Grassmann manifold and interpolates within its tangent space, preserving essential properties such as orthogonality. To provide a broader link for the reader, local \ac{PMOR} can in fact be seen as a special case of the global modal parameterization concept~\cite{Bruels2007}, which parameterizes the reduced order basis as a function of the system's own configuration rather than external parameters.

Local \ac{PMOR} approaches have received notable attention within the domain of flexible \acp{MBS}. Fischer and Eberhard~\cite{Fischer2015} apply the matrix interpolation of Panzer et al. and the Grassmann-manifold interpolation of Amsallem and Farhat to \acp{MBS} with moving loads to arrive at parameter-dependent reduced equations that closely approximate the full-order deformation response. Tamarozzi et al.~\cite{Tamarozzi2014} propose an alternative for the problem of \acp{MBS} with moving loads in which the reduction basis varies continuously in time through a parametric load position, requiring additional terms in the equations of motion. Baumann and Eberhard~\cite{Baumann2017} extend local \ac{PMOR} to systems with material removal. They interpolate finite element models at different machining stages to represent intermediate configurations. Although not specifically for \acp{MBS}, Ritzert et al.~\cite{Ritzert2023} present a local \ac{PMOR} approach for static mechanical structures to make substructure reductions parameter-dependent. While these studies demonstrate the utility of local \ac{PMOR} for mechanical systems, they also share a common feature. The reduced models change with the parameters, again defeating the purpose of reduction.

In contrast, global \ac{PMOR} preserves the conceptual simplicity of classical reduction. The system is projected onto a fixed subspace, and the reduced coordinates retain their physical meaning across the entire parameter range. Despite this advantage, however, global \ac{PMOR} has received considerably less attention in the context of flexible \ac{MBD}. To the best of our knowledge, no study has systematically applied and evaluated global reduction within a constraint multibody formalism.

The present work addresses this gap by formulating and evaluating a snapshot-based global \ac{PMOR} approach for sliding beams. The beam is discretized using planar \ac{EB} finite elements, and a sliding \ac{BC} is enforced using an algebraic constraint. Principal contributions are twofold. First, we demonstrate how a global reduction basis can be integrated into the constraint multibody formalism, using a mechanically transparent yet industrial important system. Second, we provide a systematic evaluation of the approach by varying parameters of the global reduction, benchmarking the results against a reference solution obtained with the \ac{ANCF} and an established sliding joint implementation.

The remainder of this article is structured as follows. First, the addressed mechanical problem is described. Then, the methods section presents the formulation of the global basis, the enforcement of constraints within a multibody formalism, and the \ac{ANCF} benchmark model. The experiments section describes numerical simulations to evaluate the proposed approach. Finally, the paper concludes with a discussion and summary of findings.

\begin{figure}
    \centering
    \includegraphics[scale=1]{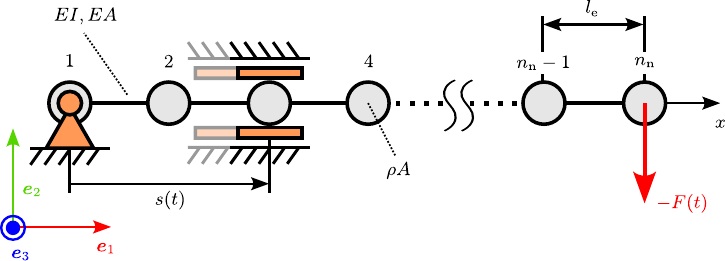}
    \caption{Schematic representation of the planar sliding beam, discretized using $n\nod$ nodes, with the discs illustrating the beam's nodes -- In the undeformed configuration, the beam axis $x$ runs parallel to $\ev_1$ and the element spanned by two adjacent nodes has a length of $l\ele$. The displacements of the first node are constrained. The parameter $s$ defines the slider position. The slider can be moved continuously, as illustrated by its transparent representation. At the last node of the beam, a force $F$ acts that is always parallel to the $\ev_2$-direction, i.e., the \emph{transverse} direction.}
    \label{fig:M_beamSchematic}
\end{figure}

\section{Problem Setup}
\label{Problem Setup}
The mechanical system under consideration, a planar flexible beam, is illustrated in Figure~\ref{fig:M_beamSchematic}. The following paragraphs, structured into three parts, review the key properties of this \ac{EB} beam.

\subsection{General Overview and Assumptions}
\label{General Overview and Assumptions}
\begin{figure}
    \centering
    \includegraphics[scale=1]{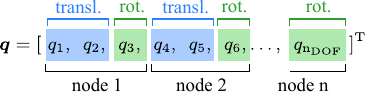}
    \caption{Structure of the coordinate vector $\qv$ -- Indexing starts at $1$ and goes up to $\nDOFs$ for the non-reduced system. Each node contributes two translational (transl.) displacements and one rotational (rot.) displacement.}
    \label{fig:M_coordinatesIndexing}
\end{figure}

The beam is discretized using $n\nod$ nodes, each carrying three \acp{DOF}: two translational displacements and one rotation. The global coordinate vector reads $\qv \in \mathbb{R}^{\nDOFs}$ with $\nDOFs=3\mult n\nod$ for the total number of \acp{DOF}. The global coordinate vector $\qv$ is ordered as shown in Figure~\ref{fig:M_coordinatesIndexing}. The beam consists of $n\ele = n\nod-1$ elements of equal reference length $l\ele$, being
\begin{equation}
    l\ele = \frac{L}{n\ele} \eqcomma
\end{equation}
where $L$ is the reference length of the whole beam.

The slider at position $s$ imposes a moving \ac{BC} along the beam axis, which constitutes the parametric feature addressed in this work. The beam is modeled using the linear planar \ac{EB} formulation, which assumes homogeneous and isotropic material properties (mass per unit length $\rho A$, bending stiffness $EI$, axial stiffness $EA$) that are constant along each element, with shear deformation and rotary inertia being neglected. By construction of the formulation,
\begin{enumerate}[label=EB\arabic*), ref=EB\arabic*, leftmargin=*]
    \item \label{ass:M_EB1} the deformation along the beam is parameterized with respect to the reference $x$-coordinate, which coincides with the $\ev_1$-coordinate, and
    \item \label{ass:M_EB2} under pure transverse loading, only transverse displacements and rotations arise. There are no axial displacements.
\end{enumerate}

The $x$-coordinate of a material point remains identical to its reference coordinate for all times $t$. Importantly, assuming the slider is located between nodes $i$ and $i+1$, its relative position $\alpha \in [0,1]$ within the element can be determined directly from the undeformed geometry as
\begin{equation}
    \label{eq:M_relativeElementPositionOfSlider}
    \alpha(s(t)) = \frac{s(t)-x_{i,\mathrm{ref}}}{l\ele} =
    \frac{s(t)-x_{i,\mathrm{ref}}}{x_{i+1,\mathrm{ref}}-x_{i,\mathrm{ref}}}
    \eqcomma
\end{equation}
where $x_{i,\mathrm{ref}}$ and $x_{i+1,\mathrm{ref}}$ denote the reference $x$-coordinates of the element nodes.

\subsection{Mass and Stiffness Matrices}
\label{Mass and Stiffness Matrices}
The element mass and stiffness matrices $\Mm\ele$ and $\Km\ele$ used in this work are taken from~\cite{Cook1989, Gavin2020}. Using these element matrices, the global $\Mm$ and $\Km$ are assembled in the standard finite element method sense, whereby shared-node \acp{DOF} receive contributions from all adjacent elements. The nodal force vector $\fv$ is 
\begin{equation}
    \label{eq:M_forceVector}
    \fv(t) = \vfoutp{0}{\dots}{F(t)}{0}\tp \eqdot
\end{equation}
Finally, the dynamic equation of the unconstrained system, i.e., the dynamic equation of the unconstrained beam, reads
\begin{equation}
    \label{eq:M_unconstrainedSystem}
    \Mm\ddot\qv(t)+\Km\qv(t)=\fv(t) \eqdot
\end{equation}

\subsection{Displacement Interpolation and Energy}
\label{Displacement Interpolation and Energy}
Taking the Hermite shape functions
\begin{equation}
\label{eq:M_shapeFunctions}
    \scalebox{0.9}{$\displaystyle
    \begin{aligned}
        N_1(\xi)&=1-\frac{\xi}{l\ele}\eqcomma &\qquad  N_2(\xi) &= 1 - 3\left(\frac{\xi}{l\ele}\right)^2 + 2\left(\frac{\xi}{l\ele}\right)^3\eqcomma &\qquad
        N_3(\xi) &= \xi - 2\frac{\xi^2}{l\ele} + \frac{\xi^3}{l\ele^2}\eqcomma \\
        N_4(\xi) &= \frac{\xi}{l\ele}\eqcomma & \qquad
        N_5(\xi) &= 3\left(\frac{\xi}{l\ele}\right)^2 - 2\left(\frac{\xi}{l\ele}\right)^3\eqcomma &\qquad
        N_6(\xi) &= -\frac{\xi^2}{l\ele} + \frac{\xi^3}{l\ele^2}\eqcomma
    \end{aligned}$}
\end{equation}
where $\xi \in [0, l\ele]$ is the element's local axial coordinate, the transverse displacement of the element spanned by nodes $i$ and $i+1$ is interpolated via
\begin{equation}
    \label{eq:M_interpolatedDisplacement}
    u_{\ev_2}(t, \xi) = \sum_{k=1}^6 N_k(\xi)\, q_{3(i-1)+k}(t) \eqdot
\end{equation}
Remember that $q_{3(i-1)+1}=0$ and $q_{3(i-1)+4}=0$ holds under pure transverse loading. In \cite{Cook1989}, a more detailed explanation of the shape functions can be found.

The total energy of the loaded \ac{EB} beam at time $t$ comprises the kinetic and strain energies, see, for example,~\cite{Shabana2013}, and the work done by the external force:
\begin{equation}
\label{eq:M_beamEnergy}
    H(t) = \underbrace{\frac{1}{2}\left( \dot\qv\tp(t) \Mm \dot\qv(t)\right)}_{\text{kinetic}} + \underbrace{\frac{1}{2}\left( \qv\tp(t) \Km \qv(t)\right)}_{\text{strain}} \underbrace{- \qv\tp(t)\fv(t)}_{\text{external}} \eqdot
\end{equation}

\begin{figure}
    \centering
    \includegraphics[scale=1]{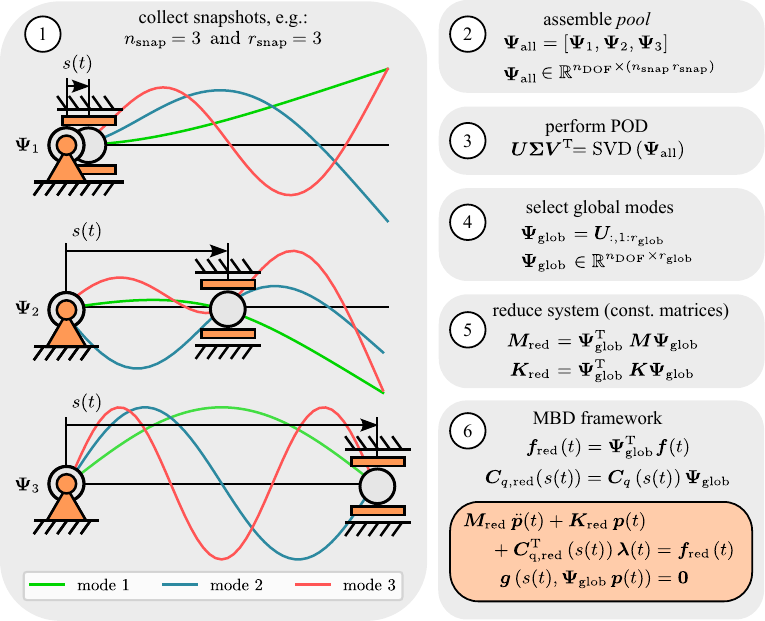}
    \caption{Overview of the proposed formulation -- Steps~(1) to~(4), covered in Section~\ref{Snapshot-Based Global Model Order Reduction}, correspond to the procedure of constructing a global reduction basis, adapted for the sliding beam problem. Steps~(5) and~(6), covered in Section~\ref{Multibody Formalism and Constraint Enforcement}, address the procedure of globally reducing the system and using it within a standard \ac{MBD} framework with algebraic constraint enforcement.}
    \label{fig:M_overview}
\end{figure}

\section{Methods}
\label{Methods}
The section is divided into three parts. The first two parts cover steps~(1) through (6) shown in Figure~\ref{fig:M_overview}. The third part deals with the \ac{ANCF}-based approach used for the benchmark model setup. Notes on the corresponding Python-based implementation\footnote{Python version 3.13; For more information on where to find and try out our implementation see the data availability statement at the end of the article.} are provided at the end of each of the three sections.

\subsection{Snapshot-Based Global Model Order Reduction}
\label{Snapshot-Based Global Model Order Reduction}
The process for the global reduction of a system is a well-documented procedure \cite{Benner2015, Panzer2010}:
\begin{enumerate}
    \item Obtain data in form of snapshots.
    \item Perform \ac{POD} on the snapshots to get a low-dimensional description of this data. \ac{POD} obtains a compact representation of data by capturing ``much of the phenomena of interest''~\cite{Chatterjee2000}.
\end{enumerate}
The following paragraphs describe the selection of snapshots, explain the computation of the snapshots, and how \ac{POD} is applied to construct the global reduction basis.

\subsubsection{Snapshot Selection}
\label{Snapshot Selection}
The $i$-th snapshot is the modal matrix $\tPsi_i$, i.e.,
\begin{equation}
    \tPsi_i = \vtretp{\tvarphi_1}{\dots}{\tvarphi_{r\snap}} \eqcomma
\end{equation}
holding the eigenmodes $\tvarphi_1$ to $\tvarphi_{r\snap}$ of the beam and with the sliding \ac{BC} being at the slider position that corresponds to the $i$-th snapshot. With eigenmodes we refer to normal modes of the mechanical system~\cite{Craig1968}. Two parameters govern the snapshot generation process: $n\snap$, the number of snapshots collected, and $r\snap$, the number of eigenmodes per snapshot.

To build intuition for this, snapshots can be viewed as photographs of the sliding beam taken with the slider at different positions. Unlike conventional photos, however, a snapshot's resolution is determined not by pixels but by eigenmodes. The more eigenmodes $r\snap$ the snapshot holds, the more detail it captures

For the $i$-th snapshot, the sliding \ac{BC} is applied at node number $\mathcal{I}_i$ when computing the modal matrix $\tPsi_i$. Since the totality of snapshots should reflect the beam characteristics across a wide range of slider positions $s$, the set of node indices is defined strategically.
\begin{equation}
    \mathcal{I} =
    \begin{cases}
        \left\{\left\lfloor \frac{n\nod}{2} \right\rfloor \right\} & \text{if } n\snap = 1 \\[10pt]
        \left\{ \left\lfloor 2 + \dfrac{(n\nod-2)\,k}{n\snap-1} \right\rfloor \;\middle|\; k = 0, \dots, n\snap-1 \right\} & \text{if } n\snap > 1
    \end{cases}
\end{equation}
The construction of $\mathcal{I}\subset\mathbb{N}$ ensures that (1) for $n\snap = 1$, the middle node is constrained with rounding being performed when there is no middle node, (2) the slider never constrains the first node with index $1$, (3) for two snapshots, the slider is at the second and last node, and (4) for more than two snapshots, the node numbers in $\mathcal{I}$ are distributed uniformly between node $2$ and node $n\nod$.

\subsubsection{Snapshot Computation}
\label{Snapshot Computation}
To compute $\tPsi_i$, the constrained \acp{DOF} are cut from the system matrices $\Mm$ and $\Km$ introduced in Section~\ref{Mass and Stiffness Matrices}. The resulting unconstrained inner~\cite{Craig1968} system matrices read
\begin{equation}
    \Mm_{\mathrm{u},i} = \Mm[\mathcal{J},\, \mathcal{J}], \qquad \Km_{\mathrm{u},i} = \Km[\mathcal{J},\, \mathcal{J}] \eqdot
\end{equation}
The set $\mathcal{J}\subset\mathbb{N}$ contains the indices of the unconstrained \acp{DOF} for snapshot $\tPsi_i$. It is constructed from the full index set $\{1, 2, \dots, \nDOFs\}$ by excluding indices $1$ and $2$, which fixes the displacements of the first node, as well as one additional index determined by the slider position used for snapshot $\tPsi_i$. For example, if the slider is located at the second node for the snapshot, index $3\mathcal{I}_i-1 = 3\cdot 2-1=5$ is additionally excluded from $\mathcal{J}$. With the resulting unconstrained system matrices, $r\snap$ eigenmodes and eigenvalues are computed. Each inner eigenmode $\tvarphi_\mathrm{u}\in\mathbb{R}^\mathrm{\nDOFs-3}$ and eigenvalue $\omega^2$ pair satisfies
\begin{equation}
\label{eq:unconstrainedEigenvaluesProblem}
    \left(-\omega^2\Mm_{\mathrm{u},i} + \Km_{\mathrm{u},i} \right)\tvarphi_\mathrm{u}=\Null \eqdot
\end{equation}
Next, a zero is inserted for the constrained \acp{DOF}, which results in the embedded and not mass-normalized modal matrix
\begin{equation}
    \tPsi^*_i=\vtretp{\tvarphi^\ast_1}{\dots}{\tvarphi^\ast_{r\snap}} \in \mathbb{R}^{\mathrm{\nDOFs} \times r\snap} \eqdot
\end{equation}
Using the global mass matrix $\Mm$, each eigenmode $j$ in $\tPsi^\ast_i$ is mass-normalized with
\begin{equation}
    \tvarphi_j = \frac{1}{\sqrt{{\tvarphi^\ast_j}\tp\Mm\tvarphi^\ast_j}}\tvarphi_j^\ast \eqcomma
\end{equation}
finally resulting in the snapshot
\begin{equation}
    \tPsi_i=\vtretp{\tvarphi_1}{\dots}{\tvarphi_{r\snap}} \in \mathbb{R}^{\mathrm{\nDOFs} \times r\snap} \eqdot
\end{equation}
Step~(1) of Figure~\ref{fig:M_overview} illustrates three snapshots, each with three mass-normalized embedded eigenmodes. Only the transverse displacement components of the first three eigenmodes are shown. Be aware that axial and rotational components in each of the eigenmodes are also present within the snapshots.

\subsubsection{Global Basis Construction}
\label{Global Basis Construction}
After computing the snapshots, a ``pool''~\cite{Panzer2010} $\tPsi_\mathrm{all}\in\mathbb{R}^{\nDOFs\times\nr}$ is assembled by stacking the snapshots. See Step~(2) in Figure~\ref{fig:M_overview}. The pool has $\nr=n\snap\mult r\snap$ columns and constitutes the raw data, which is compressed using \ac{POD}. There are different approaches to \ac{POD}~\cite{Kerschen2005}. In the following, as suggested in~\cite{Panzer2010} for the workflow of global \ac{PMOR}, we use the \ac{SVD} of the pool, see Step~(3) in Figure~\ref{fig:M_overview}.

The \ac{SVD} decomposes the real-valued\footnote{Notation of the \ac{SVD} changes if used to decompose a complex matrix. Here, only the decomposition of a real-valued matrix is considered.} pool $\tPsi_\mathrm{all}$ into three matrices~\cite{Karpfinger2026}:
\begin{itemize}
    \item $\Um\in\mathbb{R}^{\nDOFs\times\nDOFs}$ holding the left-singular vectors as columns,
    \item $\tSigma\in\mathbb{R}^{\nDOFs\times \nr}$ with all entries being zero except of the so-called singular values on the diagonal, and
    \item $\Vm\tp\in\mathbb{R}^{\nr\times \nr}$ holding the right-singular vectors as columns.
\end{itemize}
Selecting the first $r\glob$ columns, i.e., left-singular vectors, of $\Um$ yields the global reduction basis $\tPsi\glob$. See Step~(4) in Figure~\ref{fig:M_overview}. For the present application $\nr < \nDOFs$ always holds and for the singular values
\begin{equation}
\label{eq:M_singularValueOrdering}
    \sigma_{11}\geq\sigma_{22}\geq\dots\geq\sigma_{\nr\nr}\geq0
\end{equation}
applies, where the indices denote the position of each singular value within $\tSigma$. Consequently, only the first $\nr$ left-singular vectors can be assigned a singular value. The remaining singular vectors in $\Um$ are irrelevant to the transformation performed by the \ac{SVD}. Refer to~\cite{Chatterjee2000} for a more detailed discussion on singular values. An important note arises: For the number $r\glob$ of global modes, $r\glob \leq \nr$ must hold.

\subsubsection{Implementation Notes}
\label{Snapshot Based: Implementation Notes}
The NumPy\footnote{\url{https://github.com/numpy/numpy}, version 2.4.2} and SciPy\footnote{\url{https://github.com/scipy/scipy}, version 1.17.4} packages are used to solve the eigenvalue problem of Equation~\eqref{eq:unconstrainedEigenvaluesProblem} and carry out the \ac{SVD} of $\tPsi_\mathrm{all}$. During development of the proposed method, using the dense eigensolver in \texttt{scipy.linalg.eigh} without specifying a number of requested eigenmodes showed reproducible results across a Windows and a Linux machine, which is why the full dense eigensolver is recommended over both requesting only a subset of eigenvalues and eigenvectors and using a sparse solver.

\subsection{Multibody Formalism and Constraint Enforcement}
\label{Multibody Formalism and Constraint Enforcement}
The dynamic Equation~\eqref{eq:M_unconstrainedSystem} for the unconstrained beam can be extended for the sliding beam by introducing constraints via the Lagrangian Type~1 framework. The resulting system of non-reduced \acp{DAE} is
\begin{equation}
\label{eq:M_LagrangianType1}
    \begin{aligned}
        \Mm\ddot\qv(t) + \Km\qv(t) + \Cm_\mathrm{q}\tp(s(t))\tlambda(t) = \fv(t)\\
        \gv(s(t), \qv(t)) = \Null\eqcomma
    \end{aligned}
\end{equation}
where dependency on time $t$ is noted in full. According to Section~\ref{Mass and Stiffness Matrices}, $\Mm$ and $\Km$ are the system matrices. Although the mass and stiffness matrices are in general coordinate-dependent for \acp{MBS}, for the linear problem at hand, they remain constant. Refer to standard references on \acp{MBS}~\cite{Woernle2024} and references on solvers for \acp{DAE} systems~\cite{Bruels2006}. Further,
\begin{equation}
\label{eq:M_constraintJacobian}
    \Cm_\mathrm{q}(s(t))=\frac{\partial \gv(s(t),\qv(t))}{\partial \qv(t)}
\end{equation}
is the matrix of constraint gradients, i.e., the constraint \emph{Jacobian}, and $\gv$ is the binding vector. For the sliding beam, the vector $\gv$ is linear in $\qv$ and consequently the constraint Jacobian is not coordinate dependent. The Lagrange multipliers $\tlambda$ and the coordinates $\qv$ of the beam are the unknowns. Importantly, the slider position $s$ can be found in equations~\eqref{eq:M_LagrangianType1} and~\eqref{eq:M_constraintJacobian}. The slider imposes a holonomic and rheonomic constraint. It geometrically restricts the system while being time-dependent~\cite{Woernle2024}.

\subsubsection{Algebraic Constraint Enforcement}
\label{Algebraic Constraint Enforcement}
\begin{figure}
    \centering
    \includegraphics[scale=1]{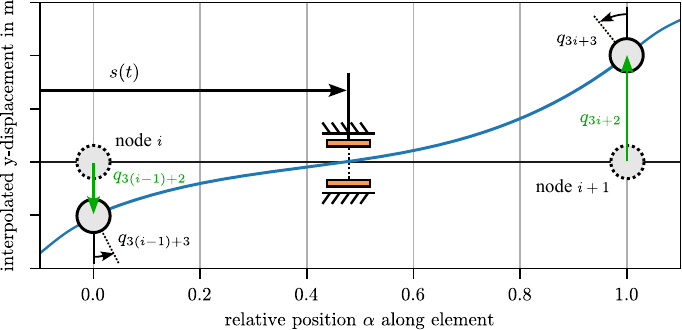}
    \caption{Schematic representation of the slider being between nodes $i$ and $i+1$ -- The slider is at position $s$, corresponding to a relative position $\alpha$ along the element. The displacement between the nodes is interpolated. At the position of the slider, the interpolated displacement is enforced to be zero using the third entry in the binding vector $\gv$.}
    \label{fig:M_yDisplacementInterpolationWithSlider}
\end{figure}

The binding vector $\gv\in\mathbb{R}^3$ is three-dimensional. Two constraints zero the displacements of the first node, i.e.,
\begin{equation}
    g_1=q_1(t) \qquad \text{and} \qquad g_2=q_2(t) \eqcomma
\end{equation}
and one constraint is added for the continuously moving slider. If the slider is between nodes $i$ and $i+1$ at time $t$, the relative position $\alpha$ is known from Equation~\eqref{eq:M_relativeElementPositionOfSlider}. Equation~\eqref{eq:M_interpolatedDisplacement} provides the interpolated $\ev_2$-displacement of the beam at this position. Therefore, $\xi$ can be replaced by $\alpha \mult l\ele$. Fully written, the third entry in the binding vector is the interpolated displacement expressed with respect to the relative slider position\footnote{The relative slider position is fully denoted as $\alpha(s(t))$. The explicit dependence of $\alpha$ on $s$ in the following is omitted.} $\alpha$:
\begin{equation}
\label{eq:M_sliderConstraint}
    g_3=N_2(\alpha)\mult q_{3(i-1)+2}(t) + N_3(\alpha)\mult q_{3(i-1)+3}(t) + N_5(\alpha)\mult q_{3i+2}(t) + N_6(\alpha)\mult q_{3i+3}(t) \eqdot
\end{equation}
Figure~\ref{fig:M_yDisplacementInterpolationWithSlider} illustrates the node displacements used within Equation~\eqref{eq:M_sliderConstraint}.

With the slider moving continously along the beam, node numbers $i$ and consequently $i+1$ change. It is advantageous that the shape functions in Equation~\eqref{eq:M_shapeFunctions} for the \ac{EB} beam are $C_1$-continuous, leading to no severe disturbances when sliding from one element to another. Since the \ac{ANCF} approach uses the same shape functions, see next section, this applies for both approaches used in our work.

Using Equation~\eqref{eq:M_constraintJacobian} the constraint Jacobian follows as
\begin{samepage}
    \begin{equation}
        \mathbf{C}_\mathrm{q}(s(t)) =
        \left[
        \begin{array}{cccccccccccc}
            1 & 0 & 0 & \multicolumn{8}{c}{\cdots} & 0 \\
            0 & 1 & 0 & \multicolumn{8}{c}{\cdots} & 0 \\
            0 & 0 & 0 & \cdots & \tikzmark{N2}N_2(\alpha) & N_3(\alpha) & 0 & N_5(\alpha) & N_6(\alpha) & 0 & \cdots & 0 \\
        \end{array}\right]\eqdot
    \end{equation}
    \begin{tikzpicture}[overlay, remember picture]
        \node[below=4mm of pic cs:N2, align=center, xshift=6mm] (label) {column $3(i-1)+2$};
        \draw[<-] ([xshift=6mm, yshift=-1mm]pic cs:N2) -- (label.north);
    \end{tikzpicture}
    \par\vspace{5mm}
\end{samepage}
\noindent If the slider moves past a node and therefore changes the associated element, the position of the $N_2 \dots N_6$ block within the matrix changes.

\subsubsection{Global Reduction}
\label{Global Reduction}
Using the global reduction basis $\tPsi\glob$ obtained via the methods explained in Section~\ref{Global Basis Construction}, it is now possible to reduce the number of unknowns by expressing the coordinates $\qv$ using modal coordinates $\pv\in\mathbb{R}^{r\glob}$ via
\begin{equation}
\label{eq:M_modalTransformation}
    \qv(t) = \tPsi\glob \pv(t) \eqdot
\end{equation}
Plugging Equation~\eqref{eq:M_modalTransformation} into the differential equations of the \acp{DAE} System~\eqref{eq:M_LagrangianType1} and multiplying $\tPsi\glob\tp$ from the left gives us the reduced system as
\begin{equation}
\label{eq:M_LagrangianType1Reduced}
    \begin{aligned}
        \underbrace{\tPsi\glob\tp \Mm \tPsi\glob}_{\Mm\red} \ddot\pv(t) + \underbrace{\tPsi\glob\tp \Km \tPsi\glob}_{\Km\red} \pv(t) +
        \underbrace{\tPsi\glob\tp \Cm_\mathrm{q}\tp(s(t))}_{\Cm_\mathrm{q, red}\tp(s(t))} \tlambda(t) = \underbrace{\tPsi\glob\tp\fv(t)}_{\fv\red(t)}\\
        \gv(s(t), \tPsi\glob\pv(t)) = \Null \eqdot
    \end{aligned}
\end{equation}
The reduced system matrices $\Mm\red$ and $\Km\red$ are computed once and remain constant throughout time. See Step~(5) in Figure~\ref{fig:M_overview}. In contrast, the reduced force vector $\fv\red$ and reduced constraint Jacobian $\Cm_\mathrm{q, red}$ can change throughout time, and they are therefore computed every time step in our numerical simulations, see Step~(6). Although not done in this work, the formulation could be made more efficient by re-computing
\begin{itemize}
    \item the reduced force vector when the external force $F$ changes, and
    \item the reduced constraint Jacobian when the slider position $s$ changes.
\end{itemize}

Due to the rules for multiplying transposed matrices~\cite{Karpfinger2026},
\begin{equation}
\label{eq:M_reducedConstraintJacobian}
    \Cm_\mathrm{q, red}\tp(s(t)) = \tPsi\glob\tp \Cm_\mathrm{q}\tp(s(t)) = 
    \left( \Cm_\mathrm{q}(s(t)) \tPsi\glob \right)\tp
\end{equation}
and further, $\Cm_\mathrm{q, red}(s(t))=\Cm_\mathrm{q}(s(t)) \tPsi\glob$ holds, which is the form used in Step~(6) of Figure~\ref{fig:M_overview}. Reduction of the constraint Jacobian for reduced-order mechanical systems is discussed in~\cite{Stadlmayr2015}. It is not clear if this approach is also applicable in the context of global reduction. This question is investigated in the following experimental section, and our findings are part of the closing discussion. Regarding the binding vector, since $\gv$ is formulated in terms of the full coordinates $\qv$, see Section~\ref{Algebraic Constraint Enforcement}, the modal coordinates are transformed back to $\qv$ using Equation~\eqref{eq:M_modalTransformation} before evaluating $\gv$ within the simulation. This is reflected in the notation used: $\gv(s(t),\tPsi\glob\pv(t))$.

After simulating the globally reduced \acp{DAE}~\eqref{eq:M_LagrangianType1Reduced}, the trajectories $\pv$ and corresponding velocities $\dot\pv$ are transformed back to $\qv$ and $\dot\qv$ in a post-processing step to get meaningful physical results.

\subsubsection{Implementation Notes}
\label{Multibody Formalism: Implementation Notes}
The multibody simulation code \EXU\footnote{\url{https://github.com/jgerstmayr/EXUDYN}, version 1.10} is used~\cite{Gerstmayr2024} for the numerical simulation of the sliding beam. Among different available items within \EXU, the Python package offers a generic \ac{ODE} object intended to simulate any possible system describable using second order \acp{ODE}. The object builds upon a generic node, which holds $r\glob$ coordinates initialized with zero initial displacements and velocities. The object is then given the reduced system matrices and a function for the time-dependent reduced force vector. A marker attached to the coordinates of the generic \ac{ODE} object makes the modal coordinates $\pv$ accessible. Using the vector-valued coordinates constraint object, the marker of the modal coordinates and a marker to \emph{ground} are connected. The binding vector $\gv$ and the reduced constraint Jacobian $\Cm_\mathrm{q, red}$ are set within this item. Both depend on the current slider position $s$, with the binding vector additionally depending on the modal coordinates $\pv$.

\subsection{Benchmark Model Setup}
\label{Benchmark Model Setup}
A benchmark model for comparison with the presented, globally reduced \ac{EB} approach is set up using well-established, nonlinear, thin beam finite elements based on the \ac{ANCF} and a sliding joint implementation. The following sections briefly describe the used \ac{ANCF} cable element and the sliding connector.

\subsubsection{Absolute Nodal Coordinate Formulation Cable Element}
\label{Absolute Nodal Coordinate Formulation Cable Element}
The \ac{ANCF} cable element is based on~\cite{Gerstmayr2008}. Each node uses two coordinates for position and two coordinates for the nodal slope vector, which is tangent to the beam element at the node. The \ac{ANCF} cable has four nodal and eight element coordinates $\qv\ele\ANCF$. As with the \ac{EB} approach, the same shape functions underlie the \ac{ANCF} approach. Arranging the shape functions from Equation~\eqref{eq:M_shapeFunctions} into the shape-function matrix $\Sm \in \mathbb{R}^{2\times 8}$ yields
\begin{equation}
    \mathbf{S}(\xi) = \begin{bmatrix} N_2(\xi)\, \One & N_3(\xi)\, \One & N_5(\xi)\, \One & N_6(\xi)\, \One \end{bmatrix} \eqcomma
\end{equation}
where $\xi\in[0,l\ele\ANCF]$ is again the element's local axial coordinate, $l\ele\ANCF$ is the length of an element, and $\One$ denotes here a $2\times 2$ identity matrix.

The physical position $\rv(\xi)$ and the according slope $\rv'(\xi)$ along the element are
\begin{equation}
    \rv(\xi)=\Sm(\xi)\mult \qv\ele\ANCF \qquad \text{and} \qquad \rv'(\xi)=\Sm'(\xi)\mult \qv\ele\ANCF \eqcomma
\end{equation}
where $\Sm'$ is the first derivative of the shape function matrix with respect to the beam's axial coordinate. The velocity is defined as
\begin{equation}
    \dot\rv(\xi) = \Sm(\xi)\mult \dot\qv\ele\ANCF\eqdot
\end{equation}
There are two conceptual characteristics of the \ac{ANCF} approach, which can be compared to assumptions~\ref{ass:M_EB1} and~\ref{ass:M_EB2} of the \ac{EB} beam:
\begin{enumerate}[label=ANCF\arabic*), ref=ANCF\arabic*, leftmargin=*]
    \item The \ac{ANCF} formulation is capable of representing large deformations. Kinematic quantities are described with respect to the element coordinate $\xi$, not the reference $x$-coordinate. Nevertheless, in the equations, $\xi$ is defined over the reference element length such that changes in the current element length do not affect stiffness.
    \item Pure transverse loading generally induces axial elongation in \ac{ANCF} beam elements. This effect is illustrated, for example, in~\cite{Gerstmayr2008} using a transversely loaded cantilever beam.
\end{enumerate}

\begin{figure}
    \centering
    \includegraphics[scale=1]{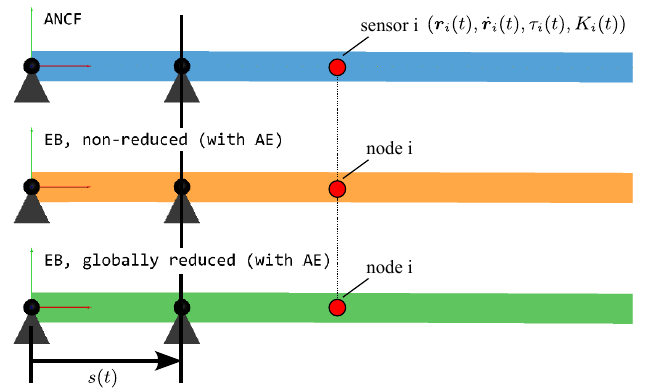}
    \caption{Simulation frame of \EXU's renderer showing the experimental setup with three approaches running in parallel -- The slider position $s$ is indicated. Sensors along the \ac{ANCF} cable coincide with the nodal positions of the \ac{EB} beams, illustrated for one sensor $i$ along the cable. The quantities measured by the sensors along the \ac{ANCF} cable are given.}
    \label{fig:E_experimentSetup}
\end{figure}

To verify the implementation of the slider using algebraic constraints within the \ac{EB}-beam-based approach, the total energy $H$ obtained via Equation~\eqref{eq:M_beamEnergy} must be compared with the total energy $H\ANCF$ obtained using the \ac{ANCF} approach. To this end, axial strain $\varepsilon$ and material curvature $K$ are needed. They are defined as
\begin{equation}
    \varepsilon(\xi) = \lVert \rv'(\xi) \rVert - 1 \eqcomma \qquad
    K(\xi) = \ev_3 \tp \frac{\rv'(\xi) \times \rv''(\xi)}{\lVert \rv'(\xi) \rVert^2} \eqdot
\end{equation}
For the \ac{ANCF} cable element, the strain at arbitrary points along the \ac{ANCF} element can become inaccurate when coupled with bending deformations~\cite{Gerstmayr2008}\footnote{See also documentation of \EXU\ regarding the \ac{ANCF} cable element: \url{https://exudyn.readthedocs.io/en/latest/docs/RST/items/ObjectANCFCable2D.html}.}. Therefore, we do not consider the axial strain contribution in the following computation of $H\ANCF$. This also improves comparability with $H$, since there axial strains are zero by construction, see Assumption~\ref{ass:M_EB2}. Using the curvature $K$, the bending moment $\tau$ becomes
\begin{equation}
\label{eq:bendingMomentANCF}
    \tau(\xi) = EI \mult K(\xi)\eqcomma
\end{equation}
where $EI$ is the bending stiffness. The total energy of the full \ac{ANCF} cable at time $t$ can then be approximated by
\begin{equation}
\label{eq:M_ANCFEnergy}
    \begin{aligned}
        H\ANCF(t) \approx &
        \underbrace{
        \frac{1}{2}\rho A
        \sum_i \frac{
        \lVert \dot{\rv}_i(t) \rVert^2 +
        \lVert \dot{\rv}_{i+1}(t) \rVert^2}{2} \Delta L
        }_{\text{kinetic}} \\
        &+
        \underbrace{
        \frac{1}{2}
        \sum_i \frac{
        \tau_i(t)K_i(t)+
        \tau_{i+1}(t)K_{i+1}(t)
        }{2}\Delta L}_{\text{strain}}
        \underbrace{-
        F(t)\,y_\mathrm{tip}\ANCF(t)}_{\text{external}}
        \eqdot
    \end{aligned}
\end{equation}
To compute $H\ANCF$, sensors indexed by $i$ are placed at equidistant positions along the \ac{ANCF} cable, spaced by $\Delta L = l\ele$, where $l\ele$ is the element length of the \ac{EB}-based approach, see Figure~\ref{fig:E_experimentSetup}. External work is computed using $y_\mathrm{tip}\ANCF$, which is the transverse tip displacement of the cable. The strain part in $H\ANCF$ can be simplified by applying $\tau_i(t)K_i(t)=EI\mult K_i^2(t)$. A detailed explanation of the terms used for computing the strain part using $K$ and $\tau$ is provided in~\cite{Bathe1996}.

The described $\Delta L$ sensor spacing further increases comparability between the \ac{ANCF} and \ac{EB} based approaches within the experiments: All compared quantities are evaluated at the same relative positions along the cable as the nodes along the \ac{EB} beam.

\subsubsection{Sliding Joint Connector}
\label{Sliding Joint Connector}
The sliding joint constrains a point, given by a marker $m_0$ to the centerline of a list of \ac{ANCF} cable elements. The sliding joint implementation follows~\cite{Pieber2022}\footnote{See also documentation of \EXU\ regarding the sliding joint: \url{https://exudyn.readthedocs.io/en/latest/docs/RST/items/ObjectJointSliding2D.html}.}. A current marker $m_1$ is used along the \ac{ANCF} cable, which itself is updated from a list of markers $[m_\mathrm{s0},\dots,m_\mathrm{sn}]$ that defines the list of connected cable elements.

\subsubsection{Implementation Notes}
\label{ANCF: Implementation Notes}
The \ac{ANCF} cable and the sliding joint connector of \EXU\ are used. A chain of $n\ele\ANCF$ cable elements is assembled. Each element uses the same parameters (mass per unit length $\rho A$, bending stiffness $EI$, axial stiffness $EA$) as the \ac{EB} elements, refer to Section~\ref{Problem Setup}. The first node is fixed in $\ev_1$- and $\ev_2$-directions using coordinate constraints against ground. The tip load $F$ is applied along $\ev_2$-direction via a coordinate load. A mass point is used to host a position marker for sliding. This mass point is constrained to ground with coordinate constraints in the $\ev_1$- and $\ev_2$-directions. The $\ev_1$-direction yields an offset that is slider position $s$. Being purely algebraic, these constraints make the added mass dynamically inactive. The simulation is commenced from zero initial
velocities and straight initial configuration without gravity.

\section{Experiments}
\label{Experiments}
Various numerical experiments are carried out. Three approaches can be considered for each, see also Figure~\ref{fig:E_experimentSetup}:
\begin{enumerate}[label=AP\arabic*), ref=AP\arabic*, leftmargin=*]
    \item \label{AP:E_ANCF} The benchmark \ac{ANCF} approach with the sliding joint, see Section~\ref{Absolute Nodal Coordinate Formulation Cable Element}.
    \item \label{AP:E_EBNR} The non-reduced \ac{EB} approach, where the non-reduced \acp{DAE}~\eqref{eq:M_LagrangianType1} is solved. In practice, $\tPsi\glob=\One\in\mathbb{R}^{\nDOFs\times\nDOFs}$ is set for this approach.
    \item \label{AP:E_EBPMR} The proposed globally reduced \ac{EB} approach, where the reduced \acp{DAE}~\eqref{eq:M_LagrangianType1Reduced} is solved.
\end{enumerate}
The beam parameters listed in the upper part of Table~\ref{tab:E_parameters} are used for all beams considered in the experiments. The parameters are chosen such that the beam exhibits large transverse deflections under moderately small loading, thereby providing a challenging benchmark for the proposed method rather than representing a specific material. The choice of parameters is part of the discussion. In the middle part of Table~\ref{tab:E_parameters}, the simulation parameters are given. Each experiment consists of a simulated time $T$, where \EXU's generalized-$\alpha$ implementation is used to simulate with a step size of $h$. See~\cite{Bruels2006} for more information on the used solver.

\begin{table}
    \centering
    \caption{Tabular representation of the parameters that are used within all of the presented experiments}
    \begin{tabular}{l|lll}
        \hline
        Parameter & Symbol & Value & Unit \\
        \hline
        \multicolumn{4}{c}{Beam Parameters}\\
        \hline
        Length & $L$ & $5$ & \si{\metre} \\
        Axial stiffness & $EA$ & $2\cdot10^7$ & \si{\newton} \\
        Bending stiffness & $EI$ & $5\cdot10^4$ & \si{\newton\square\metre} \\
        Mass per unit length & $\rho A$ & $78$ & \si{\kilogram\per\metre} \\
        \hline
        \multicolumn{4}{c}{Simulation Parameters}\\
        \hline
        Simulation End Time & $T$ & $15$ & \si{\second} \\
        Time Step Size & $h$ & $1\cdot 10^{-3}$ & \si{\second}\\
        Dynamic Solver Type & - & generalized-$\alpha$ & - \\
        \hline
        \multicolumn{4}{c}{Discretization Parameters}\\
        \hline
        \rule{0pt}{2.8ex}Number of elements of \ac{ANCF} cable & $n\ele\ANCF$ & $20$ & - \\
        Number of elements of \ac{EB} beam & $n\ele$ & $50$ & - \\
    \end{tabular}
    \label{tab:E_parameters}
\end{table}

\begin{figure}
    \centering
    \includegraphics[scale=1]{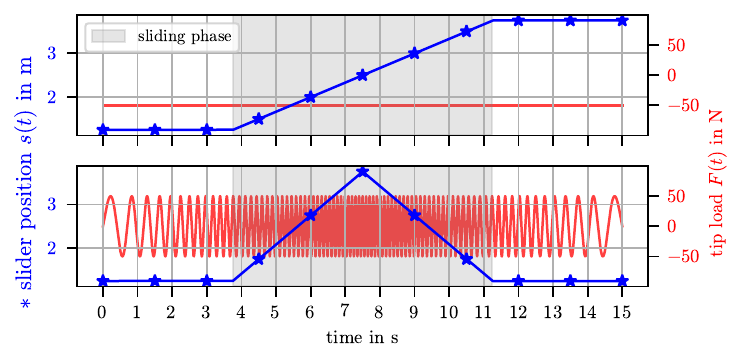}
    \caption{Slider position $s$ and tip load $F$ plotted over time for the two test cases used within the experiments. The \emph{simple} test case is shown in the upper part, while the \emph{evolved} test case is shown in the lower part. For both test cases, the sliding phase is highlighted.}
    \label{fig:E_testCases}
\end{figure}

Two test cases are considered within the experiments, see Figure~\ref{fig:E_testCases}. For the simple test case, the slider starts at $s_\mathrm{0}=L/4$ and moves within $t\in[0.25\mult T, 0.75\mult T]$ along the positive $\ev_1$-direction to $s_\mathrm{T}=3L/4$. A constant force of $F=-\SI{50}{\newton}$ is applied. For the evolved test case, the slider moves within $t\in[0.25\mult T, 0.75\mult T]$ to $s_\mathrm{T}$ and again back to $s_\mathrm{0}$. A sinusoidal force $F(t)$ of the form with varying frequency, i.e., a chirp, is applied.
\begin{equation}
    F(t) = \SI{50}{\newton}\mult \sin{\phi(t)}
\end{equation}
The phase function $\phi$ is constructed so that the force has a frequency of $f_\mathrm{0}=\SI{1}{\hertz}$ at the beginning, $f_\mathrm{T/2}=\SI{10}{\hertz}$ when the slider reaches the outmost position along the $\ev_1$-direction, and finally again a frequency of $f_\mathrm{T}=f_\mathrm{0}$. This evolved test case is intended to challenge the proposed approach by introducing a frequency-varying tip load while simultaneously varying slider position and moving it rapidly along the beam. In the implementation, a phase offset is introduced in the phase function to ensure phase continuity at the switching point between the ascending and descending chirp segments.

Two scalar error measures have been used throughout this work. The first is the normalized \ac{RMS} error measure, which is defined as follows.
\begin{equation}
\label{eq:E_NRMSE}
    e_\mathrm{NRMS}(\sv, \rv) = \min\left( \frac{1}{|\mathcal{N}|} \sum_{i \in \mathcal{N}} \frac{\sqrt{\frac{1}{J}\sum_{j=1}^{J}(s_{i,j} - r_{i,j})^2}}{\sqrt{\frac{1}{J}\sum_{j=1}^{J}(r_{i,j})^2}}, 1\right)
\end{equation}
Here, $s_{i,j}$ denotes the $i$-th component of the signal $\sv\in\mathbb{R}^{M\times J}$ under investigation at time step $j$, while $r_{i,j}$ denotes the corresponding component of the reference signal $\rv\in\mathbb{R}^{M\times J}$. In general, both signals are vector-valued with $M>1$ dimensions. For the special case of scalar-valued signals, $M=1$, Equation~\eqref{eq:E_NRMSE} also holds. Further, $J$ is the total number of time-steps within the signals, i.e., $J=T/h$. For each component, the metric computes the \ac{RMS} error between the signal and the reference over the entire time series and normalizes it by the \ac{RMS} value of the corresponding reference component. The normalized errors are then averaged over all components. To avoid divisions by reference magnitudes that are very small or zero, the used set $\mathcal{N}$ contains only those component numbers with a reference signal \ac{RMS} value greater than $1\times10^{-10}$. The normalized \ac{RMS} is capped at $1$. The second scalar error measure is the absolute error, expressed as
\begin{equation}
\label{eq:E_EABS}
    e_\mathrm{abs}(\sv, \rv) = \lVert \sv - \rv \rVert \eqdot
\end{equation}
This metric computes the norm of the difference between the signal $\sv\in\mathbb{R}^J$ and the reference $\rv\in\mathbb{R}^J$ over $J$ time steps. The equation also applies to scalars $s$ and $r$.

First, a convergence analysis determines the required number of \ac{ANCF} cable elements for the benchmark model and the number of \ac{EB} beam elements. Subsequently, selected combinations $(n\snap, r\snap)$ are examined in detail by analyzing the accuracy of the resulting global reduction basis. In a practically motivated study, we identify the optimal combination $(n\snap, r\snap)$ subject to a maximum budget of $r\glob = 20$ modal coordinates. Finally, the tip displacement and total energy are compared over time for both test cases across all three approaches~\ref{AP:E_ANCF} to~\ref{AP:E_EBPMR}.

\subsection{Convergence Analysis}
\label{Convergence Analysis}
In this experiment, the number of elements used for the \ac{ANCF} cable and the \ac{EB} beam is determined as a starting point for further experiments. Simulations are set up for approaches~\ref{AP:E_ANCF} and~\ref{AP:E_EBNR}. Successively, the number of used elements of the according formulation is increased and the sliding beams are simulated under the evolved test case. For the \ac{ANCF} approach,
\begin{equation}
    n\ele\ANCF = \{2, 4, 6, \dots,44\}
\end{equation}
is tested. For the \ac{EB} approach,
\begin{equation}
    n\ele = \{2, 4, 6, \dots,60\}
\end{equation}
is set. After each simulation, the absolute error in transverse tip displacement is computed according to Equation~\eqref{eq:E_EABS}. For each approach, the solution associated with the finest discretization serves as the reference against which all coarser discretizations are compared.

\begin{figure}
    \centering
    \includegraphics[scale=1]{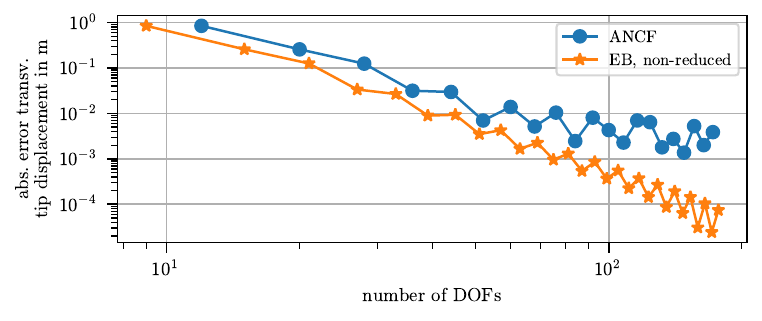}
    \caption{The absolute error of the tip displacement of the beams when using the \ac{ANCF} approach and the non-reduced \ac{EB} approach is drawn over the number of \acp{DOF} of the corresponding approach. Both axes are on a logarithmic scale.}
    \label{fig:E_convergenceAnalysis}
\end{figure}

The results obtained are shown in Figure~\ref{fig:E_convergenceAnalysis}. The number of \acp{DOF} is shown for better comparison instead of the number of elements. For both approaches, the error decreases up to approximately $52$ \acp{DOF}. For the \ac{ANCF} approach, a plateau is reached and there is no further convergence. The error with the \ac{EB} approach, however, continues to decrease. For the following experiments, $n\ele\ANCF=20$ and $n\ele=50$ are set, substantiated by the convergence analysis. See also lower part of Table~\ref{tab:E_parameters}.

\subsection{Selected Combinations Investigation}
\label{Selected Combinations Investigation}
In this experiment, three combinations of $(n\snap, r\snap)$ used to build the global reduction basis $\tPsi\glob$ according to Section~\ref{Global Basis Construction} are investigated in detail, focusing on the errors introduced by the resulting reduction bases. Two things should be remembered before conducting the experiment. First, regarding the interpretation of $n\snap$ and $r\snap$: To continue the analogy, $n\snap$ determines how many photographs of the beam with different slider positions are taken, while $r\snap$ is the resolution of each photograph, i.e., the number of eigenmodes considered within each snapshot. Second, regarding the number of global modes $r\glob$: This quantity is user-defined yet bounded by $\nr = n\snap \mult r\snap$ as explained when discussing Equation~\eqref{eq:M_singularValueOrdering}.

\begin{figure}
    \centering
    \includegraphics[scale=1]{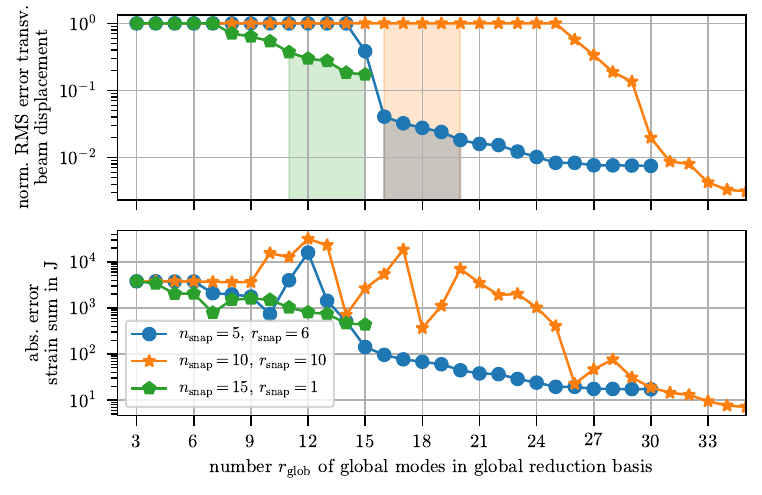}
    \caption{(Top) For three selected combinations of $(n\snap,r\snap)$, the normalized \ac{RMS} error of the transverse beam displacement is plotted when using different numbers $r\glob$ of global modes in the global reduction basis. (Bottom) The absolute error of the strain sum is plotted. In both cases, the errors are plotted on a logarithmic scale.}
    \label{fig:E_selectedCombosError}
\end{figure}

In Figure~\ref{fig:E_selectedCombosError}, three combinations are shown with two types of errors. To compute the errors, one reference trajectory for the beam coordinates is computed using the non-reduced \ac{EB} beam, i.e., Approach~\ref{AP:E_EBNR}. Afterwards, the globally reduced \ac{EB} beam is simulated, where the number of global modes $r\glob$ is successively increased for each of the combinations, starting at $r\glob=3$ up to the corresponding maximum amount. After simulation, the resulting modal coordinates are mapped back to the full coordinates. For each combination and each number $r\glob$ of global modes,
\begin{itemize}
    \item the normalized \ac{RMS} error between the reference and reduced approaches according to Equation~\eqref{eq:E_NRMSE} is computed for the $\ev_2$-displacement coordinates along the \emph{whole} beams, and
    \item the absolute error between the reference and reduced approaches according to Equation~\eqref{eq:E_EABS} is computed for the time-accumulated strain energy, where the strain energy for each time-step is the strain part in Equation~\eqref{eq:M_beamEnergy}.
\end{itemize}
For all investigated combinations, the normalized error remains at its maximum below $7$ global modes, after which the behavior diverges. For the $(15,1)$ combination, the error plateaus at approximately $0.16=\SI{16}{\percent}$ with a remaining strain energy error of \SI{600}{\joule}. The $(5,6)$ combination shows a rapid drop beyond $15$ global modes, which reaches approximately \SI{0.8}{\percent}, with the strain energy error decreasing consistently. The $(10,10)$ combination maintains its maximum error until $25$ global modes, after which both errors fall below those of the $(5,6)$ combination. The strain energy error exhibits oscillations and a spike before its descent.

\begin{figure}
    \centering
    \includegraphics[scale=1]{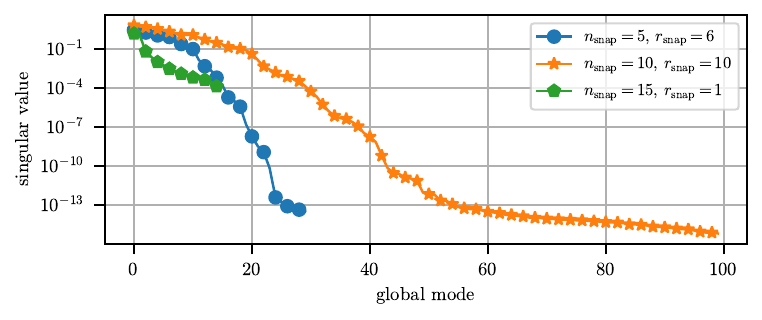}
    \caption{Singular value of the left-singular vectors when using different combinations of $(n\snap,r\snap)$ -- The singular value is plotted on a logarithmic scale.}
    \label{fig:E_selectedCombosSingularValues}
\end{figure}

Figure~\ref{fig:E_selectedCombosSingularValues} shows the singular values. Each corresponds to one left-singular vector within $\Um$, see also Section~\ref{Global Basis Construction}. The singular value descriptions reflect the ``relative importance'' of the global modes~\cite{Panzer2010}. With the combinations $(5,6)$ and $(15,1)$, the singular values decrease rapidly when considering higher global modes. This indicates that the first few modes capture most of the variance. With the combination $(10,10)$, the singular values remain comparatively high for a large number of global modes, suggesting that importance is distributed more evenly across the first global modes.

\subsection{Optimal Combinations for Maximum Budget}
\label{Optimal Combinations for Maximum Budget}
A practical engineering constraint is considered in this experiment: the maximum budget on the number of modal coordinates arising, for example, from limited or time-constrained computational resources. The objective is to find the optimal combination $(n\snap,r\snap)$ when having a maximum amount of $r\glob=20$ modal coordinates available.

\begin{figure}
    \centering
    \includegraphics[scale=1]{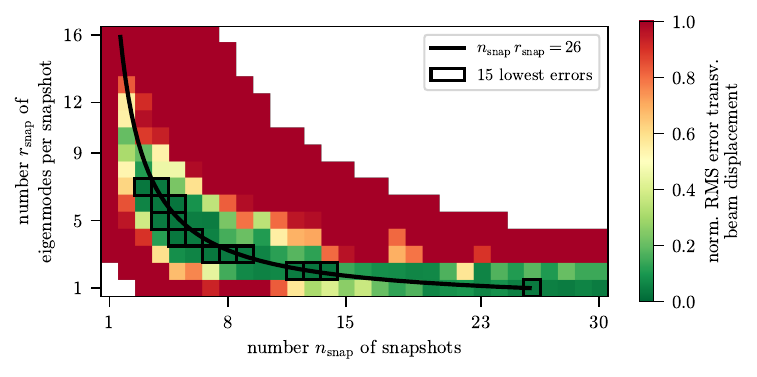}
    \caption{Normalized \ac{RMS} error of the transverse beam displacement shown for different combinations of $(n\snap,r\snap)$ if only a maximum of $20$ global modes can be used -- Each tile of the error map shows the average of maximum $5$ different investigated numbers of global modes, depending how many global modes are possible with the selected combination. In the error map, the $15$ lowest errors are also marked using black squares. The curve $n\snap\mult r\snap=26$ is shown. White areas were not investigated.}
    \label{fig:E_reductionError}
\end{figure}

To investigate this, all combinations arising from
\begin{equation}
    n\snap \in \{1, 2, \dots, 30\} \qquad \text{and} \qquad r\snap \in \{1, 2, \dots, 16\}
\end{equation}
were considered, excluding those yielding $\nr = n\snap \mult r\snap < 3$ or $\nr > 120$. For each investigated combination, the number of global modes is set to each value in
\begin{equation}
\label{eq:E_globalModesInvestigated}
    r\glob \in \{16, 17, 18, 19, 20\},
\end{equation}
if $k\geq20$. Otherwise, the highest $5$ possible numbers of global modes are used but never less than $3$. Consequently, at most five simulations were carried out per combination, each with a different number of global modes. The normalized \ac{RMS} error for transverse beam displacement is then computed for each simulation and averaged. This is illustrated in the upper part of Figure~\ref{fig:E_selectedCombosError}. For the $(5,6)$ and $(10,10)$ combinations, the full set of global modes from Equation~\eqref{eq:E_globalModesInvestigated} is available, and the corresponding normalized \ac{RMS} errors, highlighted by the shaded regions, are averaged. For $(15,1)$, only the $5$ highest numbers of global modes starting from $\nr - 4$ up to $\nr = 15$ are used. Averaging over a range of global modes avoids artificially favoring or disadvantaging any combination, because some combinations exhibit minor fluctuations in error with varying $r\glob$, a phenomenon known as the filtering effect~\cite{Benner2015}. The filtering effect is clearly visible in the strain energy error of combination $(10,10)$ at around $28$ global modes, remember Figure~\ref{fig:E_selectedCombosError}.

Applying this procedure to all combinations yields the error map shown in Figure~\ref{fig:E_reductionError}. Combinations with many snapshots and many eigenmodes per snapshot perform poorly under the maximum budget of $r\glob=20$ global modes. Accuracy increases for combinations whose product $n\snap \cdot r\snap$ is around $26$, particularly when dominated by the number of snapshots. The $15$ lowest errors cluster in a region of moderate $n\snap$ extending toward higher snapshot counts with few eigenmodes per snapshot. The optimal combination under the limited budget is $(5,6)$ with $r\glob=20$ global modes. This is a coordinate reduction from $153$ to $20$, being \SI{87}{\percent}.

\subsection{Tip Displacement, Energy, and Accuracy}
\label{Tip Displacemnent, Energy, and Accuracy}
In this experiment, the combination $(5,6)$ that proved most accurate when using $20$ global modes is used. All three approaches~\ref{AP:E_ANCF} to~\ref{AP:E_EBPMR} are simulated in parallel, successively for the simple and evolved test case. Transverse tip displacement, beam displacement, as well as the energy of the three approaches, computed either via Equation~\eqref{eq:M_beamEnergy} or Equation~\eqref{eq:M_ANCFEnergy}, is recorded over time.

\begin{figure}
    \centering
    \includegraphics[scale=1]{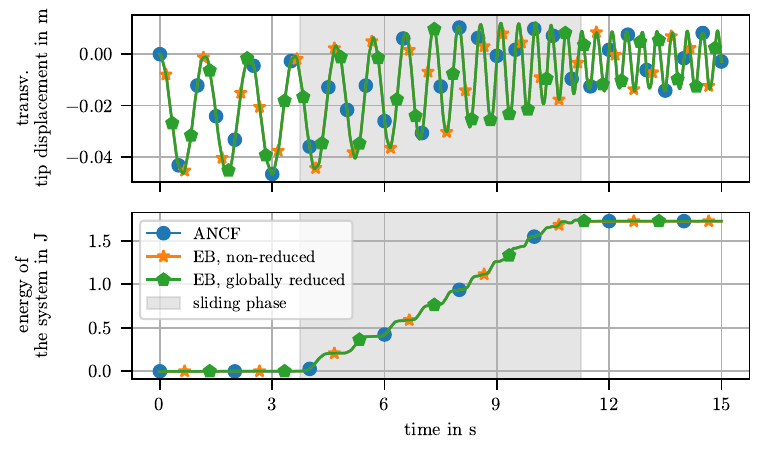}
    \caption{The beams are simulated using the \emph{simple} test case. For the proposed globally reduced \ac{EB} approach, ($n\snap=5$, $r\snap=6$, $r\glob=20$) applies. (Top) -- Tip displacement of the beam plotted over time. (Bottom) -- Energy of the three approaches plotted over time. The gray area indicates the duration of slider movement.}
    \label{fig:E_energyTipDisplacementSimple}
\end{figure}

Figure~\ref{fig:E_energyTipDisplacementSimple} shows the results for the simple test case. As expected, under constant loading, without slider movement and damping, the beam tips begin to oscillate sinusoidally with constant amplitude. The energy of the beams remain zero. As soon as the slider begins to move from the left to the right, amplitude decreases and the mean value of the tip displacement increases, as does the energy. The proposed constraint enforcement for the \ac{EB} approach, see Section~\ref{Algebraic Constraint Enforcement}, matches, with no visible differences, the sliding joint implementation for the \ac{ANCF} cable with respect to the energy. After sliding stops, the energy remains constant at a level of approximately \SI{1.75}{\joule} in all three cases. The amplitude of the tip displacement is smaller, and the mean value is closer to \SI{0}{\metre}, while at the same time the frequency of the sinusoidal movement increases.

\begin{figure}
    \centering
    \includegraphics[scale=1]{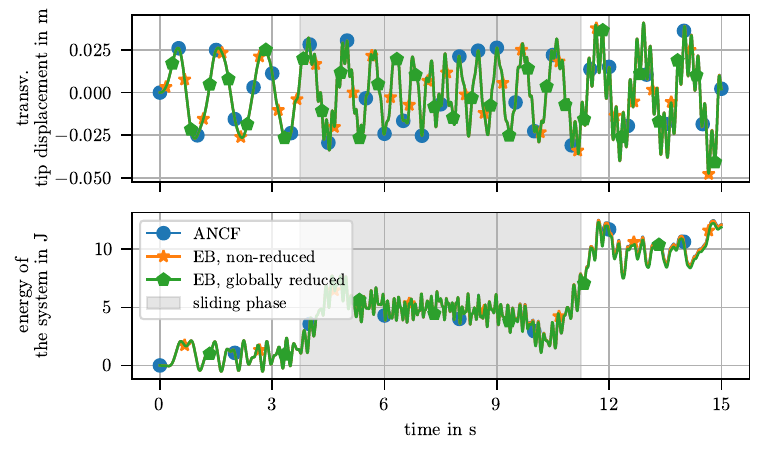}
    \caption{The beams are simulated using the \emph{evolved} test case. For the proposed globally reduced \ac{EB} approach, ($n\snap=5$, $r\snap=6$, $r\glob=20$) applies. (Top) -- Tip displacement of the beam plotted over time. (Bottom) -- Energy of the three approaches plotted over time. The gray area indicates the duration of slider movement.}
    \label{fig:E_energyTipDisplacementEvolved}
\end{figure}

Figure~\ref{fig:E_energyTipDisplacementEvolved} gives the same plots, but for the evolved test case. Although many phenomena occur simultaneously, the tip positions and the energy of the systems align visually between all three approaches. Note the high frequency parts in both, the tip displacements, and the energy.

\begin{table}
    \centering
    \caption{Normalized \ac{RMS} displacement errors for both test cases, computed for the transverse tip and beam displacement according to Equation~\eqref{eq:E_NRMSE} -- For the globally reduced \ac{EB} approach ($n\snap=5$, $r\snap=6$, $r\glob=20$) applies.}
    \label{tab:E_displacementErrors}
    \begin{tabular}{l|cc}
        \toprule
         & \textbf{Tip displacement} & \textbf{Beam displacement} \\
        \midrule
        \multicolumn{3}{c}{\textbf{simple}} \\
        non-reduced \ac{EB} vs. \ac{ANCF} & \SI{0.07}{\percent} & \SI{0.14}{\percent} \\
        globally reduced \ac{EB} vs. \ac{ANCF} & \SI{0.16}{\percent} & \SI{0.35}{\percent} \\
        \rowcolor{gray!20}
        globally reduced \ac{EB} vs. non-reduced \ac{EB} & \SI{0.20}{\percent} & \SI{0.40}{\percent} \\
        \midrule
        \multicolumn{3}{c}{\textbf{evolved}} \\
        non-reduced \ac{EB} vs. \ac{ANCF} & \SI{0.46}{\percent} & \SI{0.67}{\percent} \\
        globally reduced \ac{EB} vs. \ac{ANCF} & \SI{1.57}{\percent} & \SI{2.45}{\percent} \\
        \rowcolor{gray!20}
        globally reduced \ac{EB} vs. non-reduced \ac{EB} & \SI{1.13}{\percent} & \SI{1.82}{\percent} \\
        \bottomrule
    \end{tabular}
\end{table}

Finally, the normalized \ac{RMS} error of the transverse tip and beam displacement between the three approaches is computed according to Equation~\eqref{eq:E_NRMSE}, with results reported in Table~\ref{tab:E_displacementErrors}. The \ac{ANCF} comparisons primarily serve to confirm correct system behaviors, consistent with the energy results of Figures~\ref{fig:E_energyTipDisplacementSimple} and \ref{fig:E_energyTipDisplacementEvolved}. More importantly, the comparison between the non-reduced and globally reduced \ac{EB} formulations, see the gray rows in the table, directly quantify the errors introduced by the reduction. For the simple test case, this amounts to \SI{0.20}{\percent} for the tip and \SI{0.40}{\percent} for the beam displacement. As expected, error increases for the evolved test case, reaching \SI{1.13}{\percent} and \SI{1.82}{\percent}, respectively.

\subsection{Computational Speed}
\label{Computational Speed}
The computational speed of the three approaches~\ref{AP:E_ANCF} to~\ref{AP:E_EBPMR} is compared. Each approach alone is added to a fresh \texttt{mbs} object of \EXU, to determine solver time alone. The simulation is repeated five times for statistical reliability. All simulations were performed on a consumer laptop\footnote{Processor: 13\textsuperscript{th} Gen Intel Core i7-1370P (mobile)}.

\begin{figure}
    \centering
    \includegraphics[scale=1]{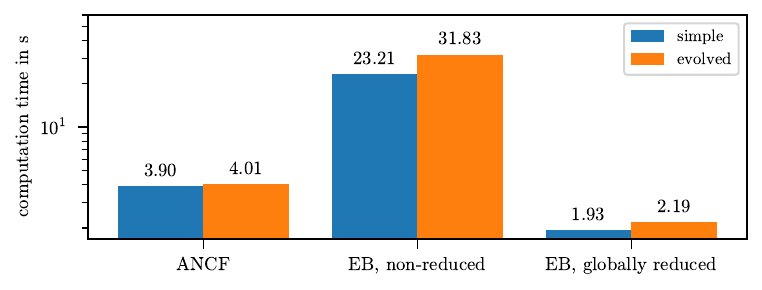}
    \caption{The computation time needed to simulate the two test cases over a time of $T=\SI{15}{\second}$ is shown for the three investigated approaches. The computation time is plotted on a logarithmic scale.}
    \label{fig:E_computationalSpeed}
\end{figure}

The results are shown in Figure~\ref{fig:E_computationalSpeed}. Although the \ac{ANCF} cable object with the sliding joint implementation needs a computation time of around \SI{4}{\second} for both test cases, computation time significantly increases by a factor of roughly $6$ to $8$ when using the \ac{EB} approach without reduction. Using the proposed global reduction basis, computation time drops by approximately \SI{90}{\percent} in both cases compared to the non-reduced \ac{EB} beam approach. Compared to the \ac{ANCF}-based approach, computation time drops by approximately \SI{50}{\percent}. This also means that, although the non-reduced \ac{EB} approach is not capable of performing real-time simulations, the reduced approach is. Remember that the reduction of computation time is the result of problem-size reduction, i.e., dropping from $153$ to $20$ coordinates, and it is accompanied by the reported loss of accuracy.

\section{Discussion}
\label{Discussion}
The discussion in the following paragraphs is structured into three parts in this order: overall experimental results are elaborated and linked to the literature, the findings regarding the snapshot parameters $(n\snap, r\snap)$ are discussed, and finally the accuracy of the presented method is assessed.

\subsection{Overall Findings}
\label{Overall Findings}
A first notable finding concerns the conditions under which the globally reduced constraint Jacobian causes the solver to fail or remain stable. Global reduction also reduces the constraint Jacobian $\Cm_\mathrm{q}$ to $\Cm_\mathrm{q,red}$ according to Equation~\eqref{eq:M_reducedConstraintJacobian}. All of the tested global reduction bases with $r\glob\geq3$ preserve solver convergence, which is not self-evident as shown in ~\cite{Stadlmayr2015}. The lower bound $r\glob\geq3$ used in the experiments is therefore not arbitrary. As shown in~\cite{Stadlmayr2017}, the solver Jacobian becomes singular when the number of constraint equations exceeds the number of modal coordinates. Since the binding vector in this formulation is always of dimension $3$, any basis with $r_\mathrm{glob} < 3$ causes solver failure.

A second finding is that the energy of the system when using the proposed algebraic constraint enforcement within the \ac{EB} approach matches, without further tuning, the energy of the established \ac{ANCF} benchmark approach with sliding joint implementation. Especially for the simple test case, the time evolution of the energy stays interpretable: although the slider moves forward, it must act against the deformed beam, i.e., the beam axis must be pushed towards the centerline. Although the Lagrange multipliers enforcing the sliding constraint within the \ac{EB} and the \ac{ANCF} sliding joint approach perform no work, the energy in the beam itself increases due to slider movement. The comparison with the established benchmark shows that for an \ac{EB} approach, the proposed constraint enforcement neither introduces dissipation nor injection of energy. To better visualize this, think of a mass point whose potential energy increases under a prescribed algebraic displacement constraint, even if the constraint itself does no work.

Another point in the discussion concerns the choice of beam parameters used in the numerical experiments, remember upper part of Table~\ref{tab:E_parameters}. Initial tests were conducted with a more engineering-relevant beam made of high-stiffness steel, and the slider constraint was enforced discretely by setting
\begin{equation}
    g_3=q_{3(i-1)+2}(t)\eqcomma
\end{equation}
where $i$ is the index of the node nearest to the slider position. Due to the high stiffness and small deformations, the energies matched closely, masking an underlying formulation error. This error only became apparent when testing a generic material with larger deformations, which motivated both the refinement to the formulation presented here and the use of the more demanding generic beam parameters given in the upper part of Table~\ref{tab:E_parameters}.

\subsection{Results regarding tested Combinations}
\label{Results regarding tested Combinations}
Recalling the analogy of $n\snap$ being the number of photographs, i.e., snapshots, and $r\snap$ being the resolution of the photographs, i.e., eigenmodes per snapshot, a high number of snapshots and eigenmodes per snapshot should yield lower displacement errors. However, when having a maximum budget of global modes, this seems not to be the case. The reason may lie in the structure of the singular value spectrum shown in Figure~\ref{fig:E_selectedCombosSingularValues}. A typical criterion for selecting the number of global modes $r\glob$, found in~\cite{Benner2015}, is
\begin{equation}
\label{eq:M_globalModeSelection}
    \frac{\sum_{i=1}^{r\glob} \sigma_i^2}{\sum_{i=1}^{\nr} \sigma_i^2} > \kappa\eqcomma
\end{equation}
where $\kappa$ is often set to $0.999$, and $\nr = n\snap \mult r\snap$ is the number of all singular values greater than zero, see Equation~\eqref{eq:M_singularValueOrdering} and the corresponding explanations. As shown in~\cite{Volkwein2011}, aiming even for a seemingly small value for $1-\kappa$ directly corresponds to a provably larger approximation error. The combination $(10, 10)$ reveals a slowly decaying singular value spectrum, so that at $n_\mathrm{glob} = 20$ the left-hand side of Inequality~\eqref{eq:M_globalModeSelection} reaches $0.999986$. In contrast, the combinations $(5, 6)$ and $(15, 1)$ achieve, within numerical accuracy, $1.0$, meaning the basis is essentially complete within the given dimension. For the $(15, 1)$ combination, although the basis is complete, the snapshots simply do not contain sufficient information about the system's behavior across the parameter space, such that accuracy remains poor regardless of the basis quality.

For the engineer, an important practical outcome follows for the problem at hand. When the budget is unlimited, it is advisable to include as much information as possible by increasing the number of snapshots and the number of eigenmodes per snapshot. When the budget is limited, however, a careful balance must be found: too little information leads to a basis that, while complete, lacks the necessary system knowledge, whereas too much information degrades the efficiency of the truncation. A tradeoff must be found and the optimum for the considered example appears to lie around $\nr = n\snap \mult r\snap = 26$. However, this is also a clear limitation of the approach: The border between good and bad working combinations is sharp and putting too much information into the reduction basis yields bad results with respect to the displacement error. The engineer must be cautious here, and a study like the one shown in the Figure~\ref{fig:E_reductionError} is recommended.

\subsection{Accuracy}
\label{Accuracy}
In the evolved test case, an error of \SI{1.82}{\percent} is observed for the beam displacement for the entire beam when comparing the non-reduced and globally reduced \ac{EB} approaches. To put these values into perspective, the maximum absolute displacement of any node in the evolved test case is \SI{47.86}{\milli\metre}, whereas the maximum absolute error at any node is \SI{0.76}{\milli\metre}. In other words, the largest error remains below \SI{1}{\milli\metre}, a length scale comparable to the thickness of a typical credit card, although the structure undergoes displacements exceeding \SI{4}{\centi\metre}. Considering the accompanying reduction in computation time of approximately \SI{90}{\percent}, this level of accuracy appears reasonable for applications such as the real-time simulation of telescopic structures.

\section{Conclusion}
\label{Conclusion}
\acresetall 
This article presented and evaluated a snapshot-based global \ac{PMOR} approach for sliding beams within a constraint multibody framework. The principal findings are summarized in five statements.

First, a global reduction basis was constructed from snapshots in the form of modal matrices, each computed with the slider at a different position along the beam. \ac{POD} via \ac{SVD} was then applied to compress the snapshot pool into a compact global basis that is valid across the entire parameter range.

Second, the global reduction basis was integrated into a constraint multibody formalism, which constitutes the core novelty of this work. Algebraic sliding constraints were formulated directly using the Hermite shape functions of the \ac{EB} beam.

Third, the proposed approach was validated against an established benchmark consisting of an \ac{ANCF} cable element combined with a sliding joint implementation. Energy comparisons confirmed that constraint enforcement neither introduces artificial energy dissipation or injection, and that the physical effects of slider movement are correctly captured.

Fourth, the influence of the snapshot parameters $(n\snap, r\snap)$ on accuracy was systematically studied for both unlimited and budget-constrained settings. Under a maximum budget of $r\glob = 20$ global modes, a tradeoff was identified. Too few snapshots leaves the basis uninformed about system behavior, while too many snapshots results in inefficient truncation.

Fifth, reducing the coordinate space from $153$ to $20$ degrees of freedom yielded a computation time reduction of approximately \SI{90}{\percent} compared to the non-reduced \ac{EB} approach, while keeping the normalized \ac{RMS} beam displacement error introduced by the global reduction below \SI{2}{\percent} even for the challenging evolved test case.

A natural next step is to embed the presented reduction approach within a floating frame of reference formulation, allowing the entire telescopic structure to undergo large rigid-body motion while the slider moves along the beam. This extension is currently under investigation and will be reported in a follow-up contribution.

\backmatter
\bmhead{Acknowledgments}
The authors would like to thank Scott Semken (LUT University, Mechanical Engineering) for his valuable help in refining the language of this article and improving its clarity.

\section*{Declarations}
\begin{itemize}
	\item \textbf{Competing interests}: Non-financial interests: Aki Mikkola is Editor-In-Chief and Johannes Gerstmayr is Associate Editor of \textit{Multibody System Dynamics}.
    \item \textbf{Funding}: The authors disclosed receipt of the following financial support for the research, authorship, and/or publication of this article: Business Finland, Virtual Material Engineering (VIIMA) project, Grant ID 7290/31/2023.
	\item \textbf{Data availability}: The Python implementation of the presented method, the source code used to generate the presented numerical results, and videos of the simulated sliding beams are available on GitHub: \url{https://github.com/SebiWeyrer/BoomPMR}. By running the script \texttt{dynamicsDemo.py}, interested readers can test the proposed approach for themselves.
    \item \textbf{Usage of Large Language Models (LLMs)}: LLMs were used only for targeted tasks such as spell checking and translation of text snippets from the authors' native language into English. The scientific content, argumentation, and writing were developed entirely by the authors.
\end{itemize}

\section*{Contributions}
SW: data curation, formal analysis, investigation, methodology, software, visualization, and writing of the original draft; JG: conceptualization, methodology, validation, supervision, and writing of the original draft; AM: conceptualization, methodology, validation, supervision, and writing of the original draft; GO: conceptualization, methodology, validation, supervision, and writing of the original draft; All authors reviewed the final manuscript.

\bibliography{literature}
\end{document}